\documentclass[letterpaper,twocolumn,10pt]{article}
\usepackage{usenix2019_v3}

\usepackage{tikz}
\usepackage{amsmath}

\usepackage{filecontents}

\usepackage{amsmath, amsfonts, amssymb, amsthm}
\usepackage{multicol}
\usepackage{multirow}
\usepackage{booktabs}
\usepackage{adjustbox}
\usepackage{subcaption}
\usepackage[ruled,linesnumbered]{algorithm2e}

\begin{document}

\date{}

\title{\Large \bf Trustworthy and Controllable Professional Knowledge Utilization in Large Language Models with TEE–GPU Execution}

\author{
{\rm Yifeng Cai}\\
Peking University
\and
{\rm Zhida An}\\
Peking University
\and
{\rm Yuhan Meng}\\
Peking University
\and
{\rm Houqian Liu}\\
Peking University
\and
{\rm Pengli Wang}\\
Peking University
\and
{\rm Hanwen Lei}\\
Peking University
\and
{\rm Yao Guo}\\
Peking University
\and
{\rm Ding Li}\\
Peking University
}

\maketitle

\thispagestyle{empty}

\subsection*{Abstract}
Future improvements in large language model (LLM) services increasingly hinge on access to high-value professional knowledge rather than more generic web data. However, the data providers of this knowledge face a skewed tradeoff between income and risk: they receive little share of downstream value yet retain copyright and privacy liability, making them reluctant to contribute their assets to LLM services. Existing techniques do not offer a trustworthy and controllable way to use professional knowledge, because they keep providers in the dark and entangle knowledge parameters with the underlying LLM backbone.

In this paper, we present PKUS, the Professional Knowledge Utilization System, which treats professional knowledge as a first-class, separable artifact. PKUS keeps the backbone model on GPUs and encodes each provider’s contribution as a compact adapter that executes only inside an attested Trusted Execution Environment (TEE). A hardware-rooted lifecycle protocol, adapter pruning, multi-provider aggregation, and split-execution scheduling together make this design practical at serving time. On SST-2, MNLI, and SQuAD with GPT-2 Large and Llama-3.2-1B, PKUS preserves model utility, matching the accuracy and F1 of full fine-tuning and plain LoRA, while achieving the lowest per-request latency with 8.1–11.9× speedup over CPU-only TEE inference and naïve CPU–GPU co-execution.
\section{Introduction}

Large language model (LLM) services~\cite{sun2024llumnix,wu2024dlora,zhong2024distserve, fu2024serverlessllm} have so far been fueled by large crawls of public web data, but they are reaching diminishing returns. Most accessible and useful web text has already been collected, while the remaining content is noisy, repetitive, or locked behind copyright and platform restrictions~\cite{zhou2024larger,villalobos2022will,longpre2024consent}. As model capacity grows, additional web-scale data yields only marginal improvements~\cite{zhou2024larger}, and scaling laws suggest that simply scraping more of the open web will not sustain future gains. The next generation of LLM services must therefore look beyond public data for meaningful improvements.

Consequently, the next important gains in LLM capability come from \emph{professional knowledge}~\cite{hu2024characterization,zhang2025way,tan2025scalable}: financial reasoning logic~\cite{yu2024fincon,liu2023fingpt}, legal procedures and interpretations~\cite{li2025legalagentbench,kim2025legisflow}, industrial diagnostics and control workflows~\cite{fan2023fate,xia2024unlocking}, and enterprise-specific operating playbooks~\cite{kim2025llms,chen2025ipdb}. This knowledge encodes structured rules, domain heuristics and operational constraints essential for accuracy, compliance and safety~\cite{song2025injecting}. They represent intellectual property and competitive advantage~\cite{tan2025scalable,10.1145/3669940.3707224,kandpal2025position}, not just another training set.

However, acquiring and utilizing professional knowledge is challenging because the income-risk profile is fundamentally imbalanced. Once a data provider's shared professional knowledge is trained into a model, existing training pipelines give the provider neither revenue share nor continued control~\cite{qureshi2025exploring,bourtoule2021machine}. If the integrated model later leaks proprietary logic, regulated records, or customer-specific information, the provider bears legal, financial, and reputational consequences~\cite{freeman2024exploring, grynbaum2023times}. LLM operators benefit from the opposite position: they monetize downstream usage while externalizing compliance and privacy risk to providers. Under this asymmetric income–risk tradeoff, providers are unwilling to contribute valuable professional knowledge~\cite{hine2024supporting,zhang2025incentivizing}.

We argue that making professional knowledge usable at scale requires restoring bargaining power to data providers. Specifically, data providers must plug their professional knowledge into LLM services in a way that is both \emph{trustworthy} and \emph{controllable}. Trustworthy use means the provider's knowledge remains confidential and is handled verifiably, even on third-party infrastructure~\cite{tan2025scalable,10.1145/3669940.3707224}. Controllable use means the provider retains lifecycle control: it can license knowledge to specific clients and later revoke or destroy it~\cite{hine2024supporting,kaur2022trustworthy}. Only when these guarantees hold can providers and LLM services engage in balanced economic exchange, where high-value knowledge can be contributed without unbounded, irreversible exposure.

Unfortunately, existing techniques fail to deliver trustworthy and controllable use because current systems are \emph{opaque} and \emph{non-separable}. On the trustworthy side, knowledge lifecycle (i.e., where it resides, how it transforms, and who obtains trained parameters) is invisible to providers. To compensate, cryptographic approaches such as secure multiparty computation (MPC)~\cite{zeng2025mpcache,lu2023bumblebee} and homomorphic encryption (HE)~\cite{deencryptedllm,moon2025thor} aim to make execution provably safe, but incur prohibitive overhead for billion-parameter models producing long outputs~\cite{hanke2024open,9352960,deencryptedllm}. They remain difficult to integrate into optimized GPU pipelines~\cite{moon2025thor}, failing practical throughput requirements. On the controllable side, mainstream adaptation methods, such as full-model fine-tuning~\cite{jiang2024megascale,yin2024lofit,deng2025hardening,labunets2025fun}, diffuse knowledge across backbone parameters~\cite{pasquini2025llmmap}. Once professional knowledge is trained into the base model, it becomes inseparable and irreversible~\cite{zhang2020dynamic}: providers cannot rent knowledge, track downstream usage, or cleanly revoke it after deployment. In short, today's mechanisms rely on heavyweight cryptography to patch an opaque lifecycle, and treat knowledge as permanently fused with the backbone, leaving no efficient path for control after sharing.

Our core insight is that transparency and separation form the foundation for trustworthy, controllable, and efficient professional knowledge use. This insight comes from two observations. First, Trusted Execution Environments (TEEs) provide hardware-rooted guarantees: models inside TEEs are invisible to platform operators, enabling auditable isolation~\cite{hashemi2021darknight,sun2020shadownet,hunt2018chiron,bai2025phantom,huang2022stamp,asvadishirehjini2022ginn,mo2020darknetz,tramer2019slalom,li2025teeslice,zhang2024no}. However, TEEs alone cannot run full LLM inference with provider-isolation at practical speed. Second, lightweight adapters such as LoRA cleanly separate updates from the backbone~\cite{he2025resource,wu2024dlora}. Importantly, professional knowledge is often compact relative to backbone parameters, making adapters expressive enough to encode it without modifying the entire LLM. Running adapters inside TEEs while keeping the backbone on GPUs establishes a design point preserving security, shareability, and revocability while retaining GPU performance.

To implement our insight, we implement the \underline{P}rofessional \underline{K}nowledge \underline{U}tilization \underline{S}ystem (PKUS), providing trustworthy protection and controllable capability. PKUS ensures trustworthy protection by executing professional knowledge inside TEEs, guaranteeing proprietary adapters remain visible only to their owners with auditable operations. PKUS ensures controllable capability by structurally separating adapters from the backbone. Thus, knowledge can be loaded, shared, suspended, or destroyed without affecting LLM backbones.

Developing PKUS requires addressing four technical challenges. First, data providers must construct a compact and enclave-suitable representation of their professional knowledge. PKUS introduces \textbf{EdgePrune}, a data provider-side training technique with a pruning mechanism that extracts a minimal adapter suitable for secure execution. Second, TEEs protect computation only after model parameters enter them, leaving the ingress-egress path and the lifecycle of per-data provider knowledge enclaves unprotected. Without a secure way to control the lifecycle of these enclaves, confidential knowledge would remain exposed. PKUS resolves this through \textbf{AegisProto}, a hardware-rooted protocol that authenticates data providers, enforces policy constraints, and provides auditable confidentiality for onboarding and managing protected adapters. Third, using professional knowledge from multiple data providers cannot rely on running multiple LLMs, which would introduce prohibitive latency. An efficient aggregation mechanism is required to support multi-party workflows. PKUS provides \textbf{AlignAgg}, a structure-aligned aggregation method that integrates protected adapters without duplicating LLM inference. Fourth, split execution introduces significant runtime overhead because GPU decoding and enclave execution require cross-boundary communication. Thus, PKUS introduces \textbf{SwiftSched}, a runtime scheduler that overlaps GPU and enclave computation, reduces boundary interactions, and enables practical low-latency inference even with many protected enclaves.

We implemented PKUS with 4,000+ lines of Python and 1,000+ lines of C++. We evaluate PKUS on two representative LLMs with three datasets. Our results show PKUS preserves model utility while maintaining security guarantees across multiple benchmarks, matching the accuracy and F1 scores of full fine-tuning and plain LoRA. For performance, PKUS achieves 8.1-11.9× speedup over CPU-only TEE execution and trivial CPU-GPU co-execution. 

Finally, we summarize our contributions as follows:

$\bullet$ We identify the core challenge in utilizing professional knowledge for LLM services as providing trustworthy protection and controllable capability without sacrificing scale or performance.

$\bullet$ We introduce the insight that separation and transparency provide the foundational ability for professional knowledge utilization in LLM.

$\bullet$ We implement this insight in PKUS, a split-execution system that runs data providers' adapters inside TEEs while keeping the backbone LLM on GPUs.

$\bullet$ We evaluate PKUS on two representative LLMs and three benchmarks, showing it matches high utility while reducing latency by 8.1–11.9× compared to baselines.

\section{Background and Motivation}

\subsection{Professional Knowledge in LLM Services}

As large public web corpora reach diminishing returns~\cite{zhou2024larger,villalobos2022will,longpre2024consent}, next-generation LLM services~\cite{sun2024llumnix,wu2024dlora,zhong2024distserve,fu2024serverlessllm} must derive performance gains from high-quality \emph{professional knowledge}~\cite{hu2024characterization,zhang2025way,tan2025scalable}. Professional knowledge denotes domain-specific logic and workflows built over years of practice and regulation, such as financial reasoning and risk-analysis procedures~\cite{yu2024fincon,liu2023fingpt}, legal workflows and decision patterns~\cite{li2025legalagentbench,kim2025legisflow}, industrial diagnostics and control strategies~\cite{fan2023fate,xia2024unlocking}, and enterprise-specific operational processes and playbooks~\cite{kim2025llms,chen2025ipdb}. The ability to incorporate professional knowledge becomes a key determinant of LLM capability and domain reliability~\cite{hu2024characterization,zhang2025way,zhou2024larger}.

For data providers, professional knowledge is among their most sensitive intellectual property. It captures proprietary business logic, competitive strategies, historical decision traces, and regulated operational rules~\cite{tan2025scalable,10.1145/3669940.3707224,kandpal2025position}. From their perspective, allowing an LLM service to use professional knowledge creates an asymmetric income–risk tradeoff: once absorbed into a model, prevailing practices give the provider little revenue participation~\cite{qureshi2025exploring,bourtoule2021machine}, yet the provider remains exposed to copyright, privacy, and compliance risks if that knowledge is memorized, reproduced, or misused. Recent training-data litigation, such as The New York Times' copyright suit against OpenAI and Microsoft over use and reproduction of Times articles in LLM outputs~\cite{freeman2024exploring, grynbaum2023times}, highlights this imbalance: model operators monetize services built on proprietary content, while data producers bear the burden of enforcing their rights. Under such conditions, rational data providers hesitate to contribute their highest-value professional knowledge~\cite{hine2024supporting,zhang2025incentivizing}.

This imbalance motivates treating professional knowledge as protected intellectual property that demands both \emph{trustworthy} and \emph{controllable} use in LLM services. Trustworthy use requires that a provider's professional knowledge remain confidential and be handled in ways that can be verified even on untrusted infrastructure~\cite{tan2025scalable,10.1145/3669940.3707224}. Controllable use requires that the provider retain lifecycle agency: it should decide which tenants or workloads may leverage its knowledge, under what policies, and for how long; license knowledge without disclosing raw parameters; and later audit or revoke that knowledge~\cite{hine2024supporting,kaur2022trustworthy}. Data providers need system-level mechanisms that give them credible leverage when interacting with LLM service operators, so that contributing professional knowledge does not mean permanently giving it away.

\subsection{Limitations of Existing Approaches}

Existing approaches do not meet the requirements for trustworthy and controllable use of professional knowledge in LLM services. Model fine-tuning makes professional knowledge inseparable from the base model and effectively irreversible once deployed~\cite{pasquini2025llmmap, zhang2020dynamic}, while cryptographic and TEE-based approaches struggle to scale or provide per-provider lifecycle control.

\textbf{Limitations of cryptographic approaches.}
Cryptographic computation~\cite{9352960,deencryptedllm,zeng2025mpcache} aims to compensate for opaque execution by making every operation provably safe, but resulting protocols do not scale to LLM inference. Specifically, MPC~\cite{zeng2025mpcache,lu2023bumblebee} and HE~\cite{deencryptedllm,moon2025thor} provide strong theoretical guarantees, yet their computational structure is mismatched with autoregressive decoding. MPC requires many rounds of interaction whose cost grows with model size and decoding length, making token-by-token inference prohibitively slow. HE restricts available operations and incurs ciphertext expansion and expensive polynomial arithmetic, resulting in orders of magnitude performance degradation~\cite{9352960,deencryptedllm,hanke2024open}. These limitations make cryptographic secure computation impractical for billion-parameter LLMs serving high-throughput, low-latency workloads. Moreover, existing cryptographic protocols typically protect monolithic joint computation: once professional knowledge is encoded into encrypted weights or inputs, there is no natural way for an individual provider to obtain separate execution boundaries, lifecycle visibility, or fine-grained revocation without re-running costly secure protocols over the entire model.

\textbf{Limitations of TEE-based approaches.}
TEE-based execution~\cite{hashemi2021darknight,sun2020shadownet,hunt2018chiron,bai2025phantom,huang2022stamp,asvadishirehjini2022ginn,mo2020darknetz,tramer2019slalom} offers hardware-rooted confidentiality and integrity, but existing TEE architectures do not provide per-provider isolation and end-to-end lifecycle control required for professional knowledge. CPU TEEs are limited by enclave memory size, context-switch overhead, and absence of high-throughput parallelism, causing orders-of-magnitude slowdown when executing large transformer layers~\cite{li2025teeslice,zhang2024no}. Today's GPU TEE mechanisms expose a single coarse-grained trusted domain rather than many independently attested, tenant-specific enclaves~\cite{lu2025mole}. As a result, knowledge loaded into a GPU TEE cannot be cleanly separated by data provider, nor revoked or destroyed on a per-provider basis in multi-tenant services. Finally, neither CPU nor GPU TEEs provide built-in mechanisms for securing model ingress and egress, so the end-to-end lifecycle of professional knowledge, from off-platform training through deployment and sharing to revocation, remains opaque to the provider and cannot be robustly audited.

\subsection{Motivation}

The limitations above suggest our insight: transparency and separation are the foundation for trustworthy, controllable, and efficient use of professional knowledge in LLM services. 

Our first observation is that TEEs provide the right transparency boundary for handling sensitive model components~\cite{li2025teeslice,zhang2024no}. TEEs offer hardware-rooted guarantees: code and data inside enclaves are invisible to platform operators and can be remotely attested, enabling auditable isolation. However, running full LLM inference with strong per-provider isolation entirely inside TEEs is impractical: enclave memory and performance constraints make it difficult to host large backbones and many providers simultaneously at the throughput required by real deployments.

Our second observation is that professional knowledge is often high-value yet structurally compact relative to backbone parameters, making lightweight adapters expressive enough to encode it without modifying the entire model~\cite{wu2024dlora}. Thus, the professional knowledge can be factored into a distinct adapter instead of the full LLM. If we isolate and protect only these adapters while keeping the backbone outside the trusted boundary, we substantially reduce the trusted computation scope and avoid scalability barriers faced by cryptographic methods and full-TEE inference, while turning each adapter into a natural unit of accounting, sharing, and revocation for data providers.

Together, these observations reveal a practical design point aligned with our core insight. PKUS keeps professional knowledge encoded as separate adapters within TEEs, while running the backbone model on untrusted GPUs. This separation enables trustworthy protection and controllable use of utilizing professional knowledge without sacrificing performance.

\section{Overview}
PKUS enables trustworthy and controllable utilization of professional knowledge by giving each data provider a dedicated, attested, and revocable protected space inside TEEs, and by supporting secure sharing and high-performance inference across these spaces. In PKUS, each data provider locally derives the compact adapters that represent its professional knowledge, and transfers this adapter into its corresponding enclave, where it can be safely used, shared, or revoked without exposing it to the service operator or other data providers.

PKUS adopts a split execution architecture in which the backbone LLM runs on an untrusted GPU, while enclaves hold only the adapters. During inference, the GPU performs the bulk of transformer computation and invokes enclave logic only for the small portion of computation that depends on protected knowledge. This architecture preserves the throughput of GPU inference while ensuring that all sensitive logic remains inside hardware-isolated enclaves tied to the identity and policies of their owners.

\begin{figure*}
    \centering
    \includegraphics[width=0.95\linewidth]{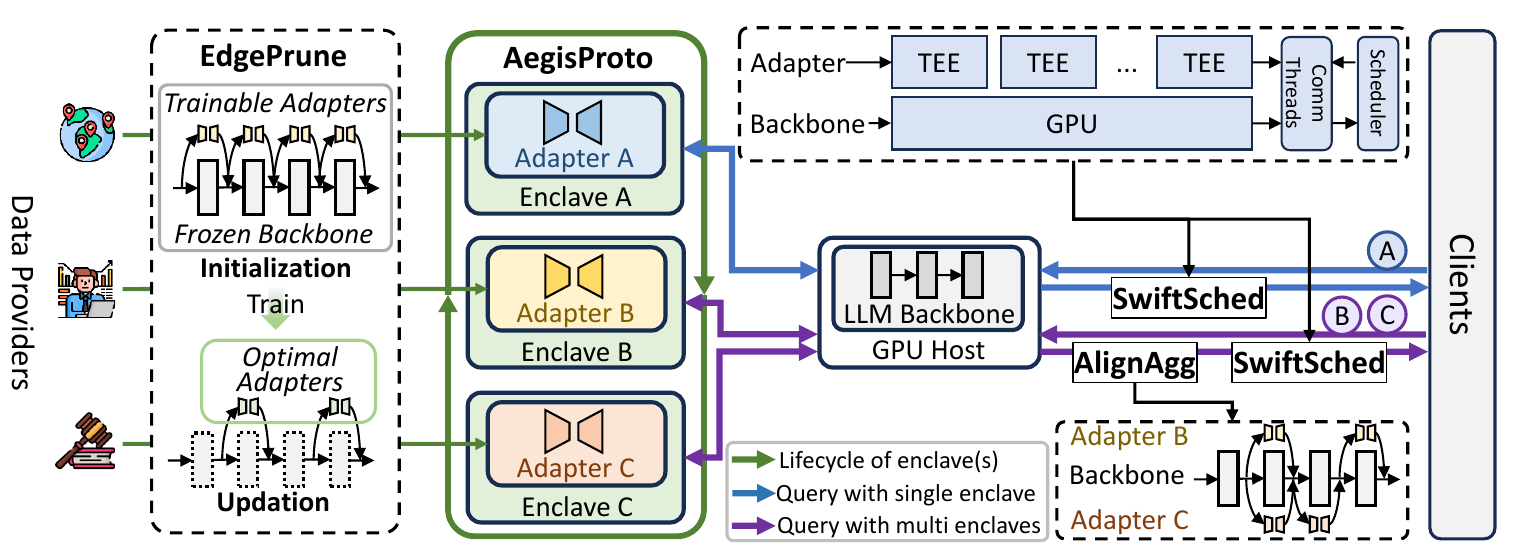}
    \caption{Overview of PKUS}
    \label{fig:overview}
\end{figure*}

To support this model, PKUS introduces four mechanisms, as shown in Figure~\ref{fig:overview}. EdgePrune helps data providers locally fine-tune their TEE-suitable professional knowledge adapters before upload. AegisProto provides a hardware-rooted protocol for securely creating, loading, and revoking per-data provider enclaves, enforcing identity binding and policy-controlled sharing. AlignAgg performs structure-aligned aggregation when multiple data providers contribute knowledge to a shared inference session, without running multiple LLM backbones. Finally, SwiftSched accelerates inference by parallelizing GPU and enclave execution, reducing cross-boundary interactions, and efficiently scheduling enclave calls. Together, these mechanisms allow PKUS to maintain end-to-end confidentiality, precise lifecycle control, and practical latency even in LLM services with professional knowledge utilization.

\section{Design}

In this section, we present the design of PKUS. We first introduce the core abstraction and execution model that structures how professional knowledge is protected and applied (Section~\ref{sec:design_abstraction}). We then describe each of our designs, including EdgePrune (Section~\ref{sec:edgeprune}), AegisProto (Section~\ref{sec:aegisproto}), AlignAgg (Section~\ref{sec:alignagg}), and SwiftSched (Section~\ref{sec:swiftsched}).

\subsection{Abstraction and Execution Model}
\label{sec:design_abstraction}

PKUS exposes a simple abstraction: each data provider owns an enclave in TEE that stores and applies its professional-knowledge adapters, while the LLM backbone runs on an untrusted GPU.

\textbf{Roles and artifacts. }
PKUS involves three roles: data providers, who hold professional knowledge; a server operator, who hosts the backbone and enclaves; and clients, who issue inference queries that may rely on one or more providers. PKUS manages the core artifacts. An \emph{adapter} is the compact representation of a provider’s professional knowledge produced by local training. An \emph{enclave} is a TEE instance bound to one provider identity that stores its adapters and executes trusted computation during inference. Enclaves are isolated from each other, from the GPU, and from the operator.

\textbf{Split execution model. }
Inference is split between untrusted and trusted components. The backbone on the GPU performs most computation. When an inference query requires specific professional knowledge, the GPU runtime issues a request to the corresponding enclave, which applies its adapters and returns a small integrity-protected response. This keeps high-bandwidth LLM computation outside TEEs while ensuring that all access to professional knowledge occurs only inside hardware-isolated enclaves.

\textbf{Knowledge lifecycle. }
Professional knowledge in PKUS follows a lifecycle of local training, secure onboarding, controlled use, and revocation. Each provider trains its adapter on local data, then securely loads it into a freshly attested enclave. Authorized clients can invoke this enclave for inference, but its parameters remain hidden from clients, the operator, and other providers. Providers can later update or revoke adapters by rotating or destroying enclaves, preserving lifecycle control over their knowledge.

\textbf{Enclave semantics. }
Enclaves export a narrow interface that applies resident adapters during inference and returns bounded outputs; they never expose internal state or merge adapters into the backbone. The server operator cannot read or clone enclave memory, and no provider gains direct access to another provider’s adapters. PKUS can coordinate multiple enclaves at inference time while maintaining strict ownership and revocation guarantees without running the entire model inside trusted hardware.

\subsection{Design of EdgePrune}
\label{sec:edgeprune}
EdgePrune enables each data provider to transform its professional knowledge into a compact adapters that can be safely loaded into an enclave. PKUS assumes that all professional knowledge originates and is trained on the data provider’s own infrastructure; EdgePrune runs entirely on the data provider side and never exposes raw training data or intermediate model updates to the server operator or other data providers.

\textbf{Design goal. }A central design goal of EdgePrune is to reduce the amount of computation that must occur inside a TEE during inference. Because the backbone model runs on an untrusted GPU while professional knowledge is applied inside per-data provider enclaves, every interaction between the GPU and an enclave introduces latency. If the enclave must evaluate a large or dense LoRA update, GPU–TEE communication becomes the dominant bottleneck. EdgePrune therefore extracts only the portions of the adaptation that meaningfully affect the data provider’s professional knowledge, producing an enclave-suitable representation that minimizes both enclave compute and cross-boundary data exchange.

The input to EdgePrune is a base LLM initialized on every transformer layer, together with the data provider’s professional data, which may include training splits used to optimize the adapters and validation data used to evaluate their impact. Through local training, the data provider produces a dense LoRA update that captures its professional knowledge but may be too large, too diffuse, or too tightly coupled to the backbone to be directly deployed inside a TEE.

The output of EdgePrune is the enclave-suitable adapters: a compact, structured representation of the data provider’s trained LoRA update that satisfies the enclave’s resource constraints and PKUS’s requirements for isolation, revocation, and multi-party aggregation. 

Concretely, let $\mathcal{S} = \{s_1, \dots, s_K\}$ denote the set of adapter sites attached to a transformer backbone, where each $s_j$ is an adapter attached to a specific linear projection. During local fine-tuning, the backbone parameters remain frozen and all task-specific knowledge is learned into the adapters. Periodically, EdgePrune computes an importance score $I(s_j)$ for each active adapter, where $I(\cdot)$ is instantiated by weight magnitude. It then ranks adapters by $I(s_j)$ and considers pruning the least important ones.

EdgePrune maintains a pruning ratio $r \in [0, r_{\max}]$ that represents the fraction of adapters that are disabled. At each pruning checkpoint, it proposes to increase $r$ by a small step $\Delta r$ and temporarily deactivates the lowest-ranked adapters to match the proposed ratio. It then evaluates the pruned model on a held-out validation set and measures the resulting metric $m(r)$ (e.g., accuracy or F1). If $m(r)$ remains above a user-specified threshold as
$m(r) \;\ge\; m(0) - \varepsilon$,
EdgePrune commits the pruning decision and updates the current ratio to $r$. Otherwise, it reverts the change and stops increasing $r$ further. In effect, EdgePrune approximates the solution to
\[
r^\star \;=\; \max_{r \in [0, r_{\max}]} \;\; r \quad \text{s.t.} \quad m(r) \ge m(0) - \varepsilon,
\]
where $m(0)$ is the validation metric of the dense adapter and $\varepsilon$ is a small tolerated drop (e.g., $0.5$--$1$ percentage point). Because the search is interleaved with fine-tuning, EdgePrune can adaptively prune more aggressively on tasks and backbones that are over-parameterized, while keeping more adapters when the task is harder.

This progressive strategy has two practical benefits. First, it automatically customizes sparsity to different model architectures and tasks without requiring per-task hyperparameter tuning of a fixed pruning schedule. Second, it produces a final adapter configuration in which only a small subset of adapters remain active, reducing both the number of LoRA parameters that must be loaded into TEEs and the number of GPU--TEE round trips during inference.

\subsection{Design of AegisProto}
\label{sec:aegisproto}
AegisProto is the protocol that manages the lifecycle of PKUS enclaves and the professional knowledge adapter they contain.
\textbf{Design goal.}
To meet the requirements of the three different roles in PKUS, AegisProto must achieve three design goals. First, it must establish a verifiable root of trust for enclaves: data providers need cryptographic evidence that their adapters will only run inside a specific, auditable runtime. Second, it must bind each enclave to an explicit ownership and policy configuration. Third, it must support secure onboarding and cleanup of adapters, ensuring that all transitions in an enclave’s lifecycle are authenticated, policy-compliant, and visible to the data providers whose adapter is affected.

\textbf{Enclave preparation.}
Enclave preparation is the process by which the PKUS operator instantiates an initialized TEE instance that can later host professional adapters. It establishes the root of trust for AegisProto: without a verifiably clean starting state, data providers cannot trust any subsequent onboarding or cleanup operations.

PKUS uses a fixed enclave runtime image containing only the components needed for AegisProto: a minimal operating environment, remote attestation support, secure channel setup, adapter management logic, and a secure cleanup routine. This image is open-source and built via a reproducible pipeline with pinned dependencies and deterministic compilation flags, so that rebuilding the same source commit yields bit-for-bit identical binaries. The resulting image is signed by PKUS and published along with its expected measurement (e.g., the TEE’s initial hash of memory and CPU state).

To prepare an enclave, the operator launches a new TEE instance using this runtime image and triggers remote attestation. The attestation quote includes the runtime’s measurement, which data providers can compare against the published value to confirm that the enclave is running the unmodified AegisProto image. Because the image is stateless and contains no persistent user data, every enclave starts from the same known-good state, avoiding cross-session contamination. The attested runtime exposes only a narrow control interface for subsequent phases of AegisProto, in order to negotiate an enclave’s ownership and policy plan, onboarding adapters, and performing cleanup when the enclave is revoked.

\textbf{Ownership and policy negotiation.}
After enclave preparation, AegisProto binds each enclave to an explicit ownership and policy plan before any professional knowledge is loaded. The plan defines (i) who owns the adapters in the enclave, (ii) which clients can invoke it. We treat the plan as a first-class object because it captures the core control semantics that PKUS must enforce throughout the enclave’s lifecycle.

For an enclave, the owner and the PKUS operator first agree on a plan that includes the base-model identifier, the owner identity, and an initial policy table. Each policy entry specifies a client identifier and optional constraints such as expiration time or maximum request volume. These entries encode the “leases” under which clients may use the owner’s knowledge: a client may be allowed to run inference for a fixed period, to participate in a particular collaboration project.

AegisProto supports dynamic policy updates to reflect evolving relationships. When a new client wishes to use an existing enclave, the client sends a signed request to the owner via the PKUS control plane. If the owner approves, it issues a signed policy update granting the requested rights, with an explicit lease duration. AegisProto records this update as part of the enclave’s policy plan. Conversely, the owner can revoke a client’s rights by issuing a signed revocation update, after which the enclave will reject further requests from that client.

Thus, the final plan hash is passed into the enclave as a launch parameter and embedded into the user-defined field of the remote attestation quote. This allows data providers and clients to verify that the enclave is running under the agreed ownership and policy configuration.

\textbf{Secure onboarding and policy updates.}
Once an enclave has been prepared and bound to a policy plan, AegisProto securely transfers adapters and policy metadata into the enclave. The main challenge is to ensure that only the intended enclave instance can install or update a data provider’s adapters and that the installed state is correctly associated with the negotiated plan.

AegisProto establishes an end-to-end encrypted channel between each owner and its enclave using an authenticated ECDH key exchange. During this exchange, the owner and the enclave derive a session key that is cryptographically bound to both the enclave’s measured runtime image and the policy-plan hash agreed in the previous phase. The owner then serializes its enclave-suitable adapters into a payload, encrypts it using an authenticated encryption scheme (i.e., AES-GCM) under the session key, and sends the resulting ciphertext through the untrusted host to the enclave.

Upon receiving the payload, the enclave reconstructs the associated data (including its own measurement and the plan hash), verifies the authentication tag, and decrypts the adapters only if all checks succeed. It then installs the adapter for that owner, without revealing it to the PKUS operator or other data providers. The enclave enforces that only the owner is allowed to perform onboarding or updates.

AegisProto uses the same secure channel mechanism to deliver policy updates. When an owner grants a new client the right to invoke or aggregate against its enclave, it prepares a signed policy-update record and sends it to the enclave over the encrypted channel. The enclave verifies the owner’s signature, updates its internal policy table, and emits an updated attestation quote that reflects the new plan hash. Clients and the PKUS operator can verify this quote to confirm that the enclave has incorporated the latest authorization decisions before routing inference requests through it. Revocation updates follow the same path and immediately take effect, preventing further use of the owner’s adapters beyond the agreed lease.

\textbf{Cleanup and revocation.}
Cleanup and revocation are the final phases in the lifecycle of an enclave and the adapters it contains. Their goal is to ensure that when an owner decides to withdraw its professional knowledge, no usable copy of the corresponding adapters remains in the TEE, and no client can continue to invoke it.

AegisProto treats cleanup as an integral part of the attested enclave runtime rather than an optional add-on. The runtime image loaded during enclave preparation includes a minimal, open-source erasure routine that is covered by the enclave’s measurement. This routine zeroes all memory regions that may contain adapters, policy tables, or intermediate inference buffers, using compiler-barrier-protected writes to prevent optimization from eliding the erasure. Because the cleanup logic is part of the measured image, data providers can verify from the attestation that the enclave they use is capable of secure erasure.

When an enclave is revoked, the owner sends a signed revocation request to the enclave via AegisProto. The enclave verifies the request, executes the cleanup routine, and then generates a final attestation quote embedding a specific revocation marker. This quote is returned to the owner and, if relevant, to affected clients. Any absence of this marker or failure to produce a valid quote indicates that cleanup did not complete successfully.

By coupling cleanup and revocation with the attested runtime and policy plan, AegisProto provides strong guarantees that  adapters in PKUS are not only protected during use but also securely withdrawn and provably erased.

\subsection{Design of AlignAgg}
\label{sec:alignagg}
AlignAgg combines professional knowledge from multiple data providers in a single inference pass without merging their adapters or replicating backbone computation. Conceptually, the input to AlignAgg is a client request together with a set of data provider enclaves that the client has selected and is authorized to use in this session. The output is the final model response produced under the aggregated effect of all selected data providers’ knowledge. AlignAgg orchestrates how the GPU backbone and the selected enclaves interact so that the backbone executes once, while each enclave contributes only its own adaptation deltas.

A naive design would instantiate a separate augmented model per data provider, run inference multiple times, and merge the final outputs. This is impractical for interactive workloads because latency grows linearly with the number of data providers. Merging weights directly inside one enclave or into the backbone would also break PKUS’s ownership and revocation semantics: once weights are mixed, individual contributions can no longer be isolated or withdrawn. AlignAgg instead exploits the structural alignment of adapters to aggregate effects, not parameters.

Specifically, for data provider $k$, EdgePrune leaves active adapters only at a subset $\mathcal{S}_k \subseteq \mathcal{S}$. At layer $s$ with input activation $x_s \in \mathbb{R}^{d_{\text{in}}}$, if $k$ has an adapter there ($s \in \mathcal{S}_k$), its enclave applies a private low-rank update $\Delta W^{(k)}_s$ and returns only the delta
$
\delta y^{(k)}_s = \Delta W^{(k)}_s x_s = A_s^{(k)} (B_s^{(k)}x_s),
$
never exposing $\Delta W^{(k)}_s$ or the adapter parameters themselves. For a client $C$, let $\mathcal{P}_C$ be the set of data providers it may use, and define the active contributors at site $s$ as
$
\mathcal{P}_C(s) \;=\; \{\, k \in \mathcal{P}_C \mid s \in \mathcal{S}_k \,\}.
$
When the backbone reaches $s$, the runtime sends $x_s$ to each enclave in $\mathcal{P}_C(s)$, collects their deltas $\{\delta y^{(k)}_s\}$, and aggregates them by simple averaging:
\[
\delta y^{(\mathrm{agg})}_s \;=\;
\begin{cases}
\frac{1}{|\mathcal{P}_C(s)|} \sum_{k \in \mathcal{P}_C(s)} \delta y^{(k)}_s, & \text{if } |\mathcal{P}_C(s)| > 0, \\
0, & \text{otherwise.}
\end{cases}
\]
Only $\delta y^{(\mathrm{agg})}_s$ is visible to the backbone; it is applied as if there were a single virtual adapter at site $s$, and decoding proceeds.

From the client’s perspective, AlignAgg turns the selected set of data providers $\mathcal{P}_C$ and a single request into one unified inference run that reflects all authorized knowledge. From the data providers’ perspective, each contribution remains encapsulated in its own enclave: revoking a data provider $k$ for client $C$ is equivalent to removing $k$ from $\mathcal{P}_C$ (and thus from all $\mathcal{P}_C(s)$), after which its deltas no longer appear in the averages and its knowledge no longer influences any outputs. No retraining, backbone change, or cross-data provider data movement is required, so multi-party aggregation remains both low-latency and fully controllable.

\subsection{Design of SwiftSched}
\label{sec:swiftsched}

SwiftSched is the runtime scheduler enabling PKUS to achieve practical low-latency inference despite split execution. It minimizes the overhead from dividing computation between the GPU backbone and CPU enclaves while preserving the strict isolation guarantees established by AegisProto.

The fundamental challenge is that split execution introduces cross-boundary interactions absent in GPU-only inference. During autoregressive decoding, each token requires the backbone to invoke enclaves at injection sites, transmit activations, wait for computation, receive deltas, and apply them before proceeding. Naive serialization makes these interactions the dominant latency source: the GPU stalls waiting for enclave responses, and enclaves idle during non-adapted layers. As data providers or injection sites increase, boundary-crossing overhead grows proportionally, threatening multi-party workflow scalability.

SwiftSched addresses this through three complementary mechanisms that overlap GPU and enclave execution, reduce boundary-crossing frequency, and efficiently schedule enclave invocations. Together, these maintain near-native GPU throughput even with many data providers.

\textbf{Pipeline parallelism between GPU and enclaves. } GPU backbone computation and enclave delta computation are largely independent within a layer. SwiftSched introduces a pipeline that decouples GPU forward progress from enclave response collection.

SwiftSched maintains a dispatch queue per layer tracking pending enclave invocations. When the GPU produces activation $x_s$, the host runtime enqueues asynchronous sends to all enclaves in $\mathcal{P}_C(s)$ and immediately continues GPU execution. A dedicated communication thread pools enclave responses as they arrive. When the backbone needs $\delta y^{(\text{agg})}_s$ to proceed, SwiftSched synchronizes with the communication thread to ensure all deltas for site $s$ have been received and aggregated. This overlap hides enclave computation latency behind GPU execution, reducing wall-clock time per token.

\textbf{Batched boundary crossing. } SwiftSched reduces boundary-crossing frequency by batching multiple injection-site requests into a single enclave invocation. The naive approach sends one message per site; for data providers with adapters spanning many sites within a layer, this creates dozens of small, latency-dominated messages per token.

SwiftSched groups all injection sites within a contiguous layer segment (e.g., all attention projections or feed-forward projections) belonging to the same data provider into a single batch. The host collects input activations $\{x_{s_1}, \ldots, x_{s_m}\}$ and transmits them in one message. The enclave processes each site, accumulates deltas $\{\delta y^{(k)}_{s_1}, \ldots, \delta y^{(k)}_{s_m}\}$, and returns them in one response. This amortizes the fixed cost of cross-boundary serialization, authentication, and context switching over multiple sites, reducing per-site overhead.

\textbf{Data provider-level scheduling and load balancing. } SwiftSched employs a data provider-level work-stealing scheduler to address enclave invocation scheduling when multiple providers contribute to the same site or have skewed sparsity patterns. For each injection site $s$, SwiftSched identifies active contributors $\mathcal{P}_C(s)$ and issues concurrent invocations. Each enclave has a dedicated worker thread managing communication and computation. When an enclave lacks pending work, its thread assists with dispatch management or prefetches next-layer activations. If an enclave's computation is slower, SwiftSched's synchronization allows other enclaves to complete without blocking, forming $\delta y^{(\text{agg})}_s$ as soon as all contributions arrive.

SwiftSched tracks per-data provider latency statistics and adjusts batching granularity dynamically: fast-responding providers receive larger batches to maximize throughput, while slower providers get smaller batches to reduce synchronization stall. This adaptive scheduling maintains balanced load even when adapters differ significantly in size, sparsity, or computational cost.

\textbf{Memory and communication optimization. }
SwiftSched minimizes cross-boundary data movement costs through low-level optimizations. Activation and delta tensors use a compact, preallocated binary format avoiding serialization overhead. Fixed-size communication buffers are reused across tokens to eliminate allocation latency. For enclaves contributing at many sites, SwiftSched prefetches next-layer activations into a staging area while current deltas are aggregated, enabling immediate computation upon the next invocation.

SwiftSched integrates with AegisProto's secure channel, applying lightweight message authentication codes (MACs) to payloads using session keys from onboarding. This preserves end-to-end confidentiality and integrity with negligible per-message cryptographic overhead compared to enclave computation and data transfer costs.

In summary, SwiftSched transforms split execution from a potential bottleneck into a practical architecture. By overlapping GPU and enclave computation, batching cross-boundary interactions, scheduling enclaves adaptively, and optimizing communication paths, SwiftSched enables PKUS to achieve inference latencies competitive with GPU-only baselines even when applying professional knowledge from multiple data providers simultaneously, critical for making trustworthy protection and controllable capability compatible with real-time enterprise LLM deployments.
\section{Evaluation}

In this section, we answer the following research questions:

$\bullet$ RQ1.How well does PKUS preserve model utility when protecting professional knowledge?

$\bullet$ RQ2. How effective is EdgePrune at reducing enclave-side computation and GPU–TEE communication?

$\bullet$ RQ3. How does AlignAgg enable efficient and accurate multi-data provider inference?

$\bullet$ RQ4. How much does SwiftSched improve the latency and scalability of PKUS?

$\bullet$ RQ5. What security guarantees does PKUS provide?

\subsection{Experiment Configurations}

\textbf{Implementation. }PKUS is implemented as a split-execution runtime that integrates on top of existing LLM toolchains. PKUS is written in about 4K lines of Python and 1K lines of C++. We extend the standard PyTorch and Transformer library with hooks for enclave calls and cross-device scheduling, and expose a minimal interface to invoke per-data provider protected enclaves.

\textbf{Dataset. }
We evaluate PKUS on three public benchmarks that approximate different forms of professional knowledge: SST-2~\cite{wang2018glue} for sentiment classification (67K train, 1.8K test movie reviews), MNLI~\cite{wang2018glue} for natural language inference (393K train, 9.8K test sentence pairs), and SQuAD 1.0~\cite{rajpurkar2016squad} for extractive QA (100K+ Wikipedia question–answer spans). Each dataset is treated as the professional knowledge of a distinct data provider and is hosted in its own enclave.

\textbf{Model and hyperparameter. }We adapt two pretrained decoder-only language models as backbone models in PKUS: GPT-2 Large~\cite{radford2019language} and Llama-3.2-1B~\cite{dubey2024llama}. GPT-2 Large is a widely used transformer model with 774M-parameter. Llama-3.2-1B is a lightweight 1B-parameter model from the Llama 3.2 family, which we use as a representative modern LLM. For both backbones, PKUS uses LoRA-style rank-8 adapters ($\alpha=16$) on the attention and feed-forward projections for fine-tuning on the professional knowledge. In all cases, we use standard supervised fine-tuning with a causal language modeling objective. The SST-2 and MNLI are tuned with a learning rate of 3e-5 for 3 epochs, and the SQuAD dataset with a learning rate of 2e-5 for 2 epochs.

\textbf{Testbed. }Our testbed involves three types of machines. The PKUS serving backend runs on a server with an AMD EPYC 7302 processor supporting AMD Secure Encrypted Virtualization (SEV), two NVIDIA RTX 3090 GPUs for high-throughput backbone inference, 64 GB of RAM, and 7 TB of NVMe SSDs for local storage of backbone checkpoints. On this machine, backbone weights are pinned in GPU memory, while per-data provider adapters reside in their corresponding enclaves on the TEE-capable CPU. The TEE, which is a confidential virtual machine (CVM), has been allocated 16 vCPUs and 16 GB of memory. Data provider-side adapter training runs on a workstation that represents each data holder’s local environment. This workstation is equipped with a commodity multi-core CPU, four NVIDIA A6000 GPUs, 128 GB of RAM, and 2 TB of storage. All adapter training is performed on this workstation, so raw training data and intermediate model states never leave the data provider’s machine. Finally, we use a MacBook Pro laptop to emulate clients that issue inference requests to the PKUS serving backend. The client machine does not perform any heavy computation and only observes attested outputs from PKUS.

\textbf{Baselines.} For each dataset–backbone pair, we consider two baselines: (1) full-model fine-tuning, which fine-tunes all parameters of GPT-2 Large or Llama-3.2-1B backbones; and (2) a LoRA baseline, which fine-tunes the full low-rank adapters without pruning. In addition, we instantiate two state-of-the-art cryptographic baselines, BumbleBee (MPC)~\cite{lu2023bumblebee} and THOR (HE)~\cite{moon2025thor}, wrapping the same model as PKUS using their implementations and recommended configurations.

\textbf{Metric. }We report \emph{accuracy} for single-label classification tasks (SST-2 and MNLI) and token-level \emph{F1} for SQuAD. All reported numbers are averaged over three runs. 

\subsection{RQ1: Utility}
\label{sec:rq1}

\begin{table}[t]
  \centering
  \caption{Task performance of non fine-tuning, full-model fine-tuning (Full-FT), plain LoRA, and PKUS on four datasets. For SST-2 and MNLI, we report accuracy (\%), and for SQuAD we report F1 score (\%).}
  \label{tab:rq1-utility}
  \begin{adjustbox}{max width=1.\columnwidth}
  \begin{tabular}{llcccc}
    \toprule
    Backbone & Task & Non-FT & Full-FT & LoRA & PKUS \\
    \midrule
    \multirow{3}{*}{GPT-2 Large}
      & SST-2 (Acc)   & 49.6 &   94.1     &   93.9   &  93.8    \\
      & MNLI (Acc)    &  32.6  &   79.8     &   85.7   &   85.7   \\
      & SQuAD (F1)    &  0.1  &   78.5     &   82.5   &   82.3   \\
    \midrule
    \multirow{3}{*}{Llama-3.2-1B}
      & SST-2 (Acc)    & 48.7 &   95.6     &   95.8   &   95.8   \\
      & MNLI (Acc)    &  32.5  &    88.8    &  89.3    &   89.2   \\
      & SQuAD (F1)    &  0.1  &   89.2     &   88.7   &    88.2  \\
    \bottomrule
  \end{tabular}
  \end{adjustbox}
\end{table}

\begin{table}[t]
  \centering
  \caption{Per-request latency (sec) for single-query inference under different deployment configurations. }
  \label{tab:rq1-latency}
  \begin{adjustbox}{max width=1.\columnwidth}
  \begin{tabular}{llcccc}
    \toprule
    Backbone & Config    & SST-2 & MNLI & SQuAD \\
    \midrule
    \multirow{5}{*}{GPT-2 Large}
      & GPU & 6.8 &   7.2   &  7.3 \\
      & TEE & 68.0 &   69.1   & 70.1 \\
      & BumbleBee (MPC) &   1535.1  & 1625.4 & 1649.0 \\
      & THOR (HE) &    4938.5   & 5229.1 & 5301.8 \\
      & PKUS & 9.9 &   11.7   &  14.4  \\
      
    \midrule
    \multirow{5}{*}{Llama-3.2-1B}
      & GPU & 4.9 &   5.3   &  5.4 \\
      & TEE & 83.9 &   85.8   & 89.0 \\
      & BumbleBee (MPC) &   2128.2    & 2301.4 & 2345.9 \\
      & THOR (HE) & 5601.1 &   6059.4   & 6172.6 \\
      & PKUS & 7.3 &   8.5   &  9.0  \\
    \bottomrule
  \end{tabular}
  \end{adjustbox}
\end{table}

PKUS preserves task utility while protecting professional knowledge inside enclaves. Table~\ref{tab:rq1-utility} compares model quality across SST-2, MNLI, and SQuAD for four configurations: a non-fine-tuned backbone (Non-FT), full-model fine-tuning (Full-FT), plain LoRA, and PKUS with enclave-executed adapters. For GPT-2 Large, PKUS attains 93.8\% / 85.7\% accuracy on SST-2 / MNLI and 82.3\% F1 on SQuAD, which is within 0.3, 0.0, and 0.2 percentage points, respectively, of the strongest baseline. For Llama-3.2-1B, PKUS reaches 95.8\% / 89.2\% accuracy and 88.2\% F1, at most 1.0 point below the best baseline on each task. Across all six tasks, PKUS never deviates by more than 1.0 absolute percentage point from the better of Full-FT and LoRA, and even slightly outperforms Full-FT on GPT-2 Large MNLI (+5.9) and SQuAD (+3.8). This confirms that confining adapters to per-data provider enclaves does not harm model utility by itself. Moreover, we attribute cases where Full-FT underperforms LoRA adapters to overfitting: updating all backbone parameters on relatively small supervised datasets can move the pre-trained representation away from a good initialization, whereas low-rank adapters act as an implicit regularizer and preserve more of the original model structure.

Meanwhile, PKUS introduces moderate end-to-end latency overhead relative to an insecure GPU-only deployment, while avoiding the dramatic slowdown of full-model TEE or cryptographic execution. Table~\ref{tab:rq1-latency} reports per-request latency for single-query inference. Averaged across SST-2, MNLI, and SQuAD, PKUS increases latency on GPT-2 Large from 7.1 sec to 12.0 sec and on Llama-3.2-1B from 5.2 sec to 8.3 sec, corresponding to roughly 69\% and 59\% overhead (1.7$\times$ and 1.6$\times$ slowdown) compared to running the entire model on the GPU. In contrast, executing the full model inside a CPU TEE is about 9.7$\times$ slower than the GPU baseline for GPT-2 Large (69.1 sec per query) and 16.6$\times$ slower for Llama-3.2-1B (86.2 sec per query). The gap is even larger compared to state-of-the-art cryptographic schemes:  BumbleBee and THOR are between roughly 226$\times$ and 1{,}143$\times$ slower than GPU-only inference, and about 134$\times$–719$\times$ slower than PKUS. Because PKUS routes only small adapter computations through enclaves, keeps the backbone on the GPU, and overlaps enclave work with decoding, it preserves most of the throughput of GPU-only inference while still providing hardware-enforced isolation for professional knowledge.

\subsection{RQ2: Effectiveness of EdgePrune}

\begin{table}[t]
  \centering
  \caption{Effect of EdgePrune on prune ratio, adapter size, and per-request latency. ``Ratio'' denotes the pruned ratio, ``Params'' is the number of adapter parameters (in millions), and ``Latency'' measures the inference time under PKUS with a single query.}
  \label{tab:rq2}
  \begin{adjustbox}{max width=1.\columnwidth}
  \begin{tabular}{llcccc}
    \toprule
    {Backbone} &
    {Task} &
    {Ratio} &
    {Params} &
    {Latency} &
    Speedup \\
    \midrule
    \multirow{3}{*}{GPT-2 Large}
      & SST-2      &    79.0\%   &     1.2 M    &  9.9 sec &   2.3$\times$ \\
      & MNLI       &    68.0\%    &     1.9 M   &  11.7 sec &   2.0$\times$  \\
      & SQuAD      &    63.0\%    &     2.1 M   &  14.4 sec &   1.7$\times$ \\
    \midrule
    \multirow{3}{*}{Llama-3.2-1B}
      & SST-2      &    80.0\%    &     0.9 M   &  7.3 sec &   2.4$\times$ \\
      & MNLI       &    66.0\%    &     1.5 M   &  8.5 sec &   2.1$\times$\\
      & SQuAD      &    61.0\%    &     1.7 M   &  9.0 sec &   1.8$\times$ \\
    \bottomrule
  \end{tabular}
  \end{adjustbox}
\end{table}

\begin{figure}[t]
    \centering
    \includegraphics[width=1.\linewidth]{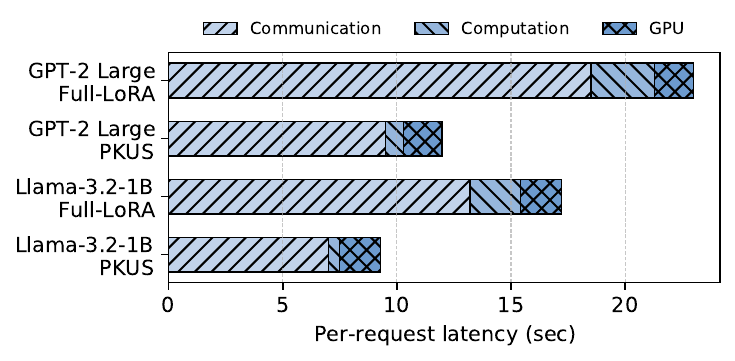}
    \vspace{-2em}
    \caption{Breakdown of per-request latency into communication, enclave computation, and GPU time.}
    \label{fig:edgeprune-breakdown}
\end{figure}

EdgePrune preserves model quality while aggressively reducing the amount of adapter computation that must run inside enclaves. For each backbone and task, we start from a dense LoRA adapter and apply EdgePrune to find the most compact configuration that stays within the same utility as the dense baseline (as reflected in Section~\ref{sec:rq1}). Table~\ref{tab:rq2} summarizes the resulting pruning ratio, parameter size of the adapter, inference latency, and speedup over the dense baseline under PKUS. On GPT-2 Large, EdgePrune prunes 63.0\%–79.0\% of the original adapter parameters (retaining about 30.0\% on average). On Llama-3.2-1B, the pruned ratio ranges from 61.0\% to 80.0\% (retaining about 31.0\% on average), again pruning away 69.0\% of parameters. Across all tasks, we do not observe systematic degradation in accuracy or F1 relative to dense LoRA adapters, indicating that a large fraction of dense adapter structure is unnecessary for these workloads.

Pruning directly translates into lower enclave-side computation and fewer GPU–TEE round-trip communications. EdgePrune reduces per-query latency under PKUS to 9.9--14.4 sec on GPT-2 Large and 7.3--9.0 sec on Llama-3.2-1B. Compared to PKUS with dense adapters, this corresponds to speedups of 1.7$\times$--2.3$\times$ on GPT-2 Large and 1.8$\times$--2.4$\times$ on Llama-3.2-1B, i.e., roughly halving adapter-related latency on average. 
To understand where these gains come from, we further break down per-request latency into communication between the GPU and enclaves, enclave-side computation, and GPU time (Figure~\ref{fig:edgeprune-breakdown}).  With EdgePrune, communication decreases by 2.0$\times$ and enclave computation by 3.6$\times$. Llama-3.2-1B shows a similar pattern: communication decreases by 1.9$\times$ and enclave computation by 4.6$\times$. Meanwhile, the GPU time remains constant. Thus, most of the speedup comes from eliminating adapter computation inside enclaves and reducing GPU--TEE computation, rather than from any change to backbone execution.
Together, these results show that EdgePrune can eliminate most adapter capacity while preserving utility.

\subsection{RQ3: Effectiveness of AlignAgg}

\begin{figure}[t]
  \centering
  \begin{subfigure}{0.48\columnwidth}
    \centering
    \includegraphics[width=\linewidth]{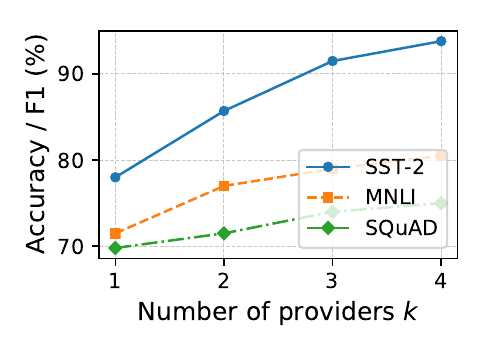}
    \vspace{-2em}
    \caption{GPT-2 Large}
    \label{fig:alignagg-gpt2}
  \end{subfigure}
  \hfill
  \begin{subfigure}{0.48\columnwidth}
    \centering
    \includegraphics[width=\linewidth]{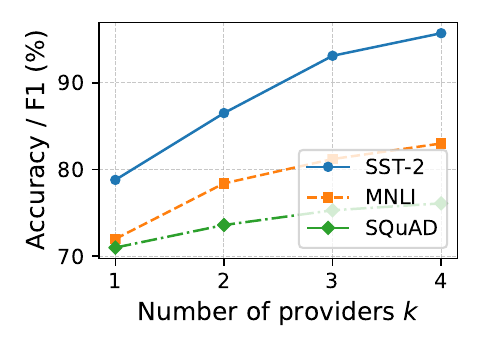}
    \vspace{-2em}
    \caption{Llama-3.2-1B}
    \label{fig:alignagg-llama}
  \end{subfigure}
  \vspace{-1em}
  \caption{Effectiveness of AlignAgg as the number of data providers $k$ increases. Results are averaged over all $\binom{4}{k}$ subsets. The metric is accuracy (\%) for SST-2 and MNLI, and token-level F1 (\%) for SQuAD.}
  \label{fig:alignagg-k}
\end{figure}

\begin{table}[t]
  \centering
  \caption{Latency (sec) with aggregated enclaves.}
  \label{tab:rq3-scale}
  \begin{adjustbox}{max width=1.\columnwidth}
  \begin{tabular}{lcccccccc}
    \toprule
    Backbone \verb|\| \#Enclaves &
    1 & 
    2 &
    4 &
    8 &
    16 &
    32  \\
    \midrule
    GPT-2 Large  &
      12.0 & 15.4 & 20.6 & 26.9 & 31.2 & 32.4         \\
    \midrule
    Llama-3.2-1B &
      8.3 & 12.4 & 16.8 & 22.2 & 26.2  & 27.7        \\
    \bottomrule
  \end{tabular}
  \end{adjustbox}
\end{table}

For each dataset (SST-2, MNLI, and SQuAD) and each backbone, we treat the PKUS configuration in Section~\ref{sec:rq1} (trained on the full dataset) as an upper bound on utility. We then randomly partition the training data into four equal shards and train one adapter per shard, keeping the backbone frozen. Each shard is treated as a logically independent data provider with its own enclave. At evaluation time, we use AlignAgg to aggregate adapters from $k$ data providers, $k \in \{1,2,3,4\}$. We report the average accuracy or F1 over all $\binom{4}{k}$ choices of $k$ data providers to smooth out the effect of a particular split in Figure~\ref{fig:alignagg-k}.

AlignAgg provides clear utility gains as more data providers are aggregated and can closely match the upper bound on simpler tasks. On SST-2, moving from a single data provider to four data providers improves GPT-2 Large from 78.0\% to 93.8\% and Llama-3.2-1B from 78.8\% to 95.7\%. With four data providers, AlignAgg exactly matches the full-data PKUS upper bound for GPT-2 Large and comes within 0.1 percentage points for Llama-3.2-1B. This suggests that the sentiment task is sufficiently redundant that different shards learn complementary but overlapping patterns, and aggregating their adapters effectively reconstructs the behavior of a model trained on all data.

On MNLI and SQuAD, AlignAgg still improves as $k$ increases but retains a moderate gap to the upper bound. For GPT-2 Large, $k=4$ reaches 80.5\% accuracy on MNLI and 75.0\% F1 on SQuAD, compared to 85.7\% and 82.3\% F1 for full-data PKUS, i.e., gaps of 5.2 and 7.3 percentage points. For Llama-3.2-1B, the corresponding values are 83.0\% vs.\ 89.2\% on MNLI and 76.1\% vs.\ 88.2\% F1 on SQuAD, gaps of 6.2 and 12.1 points. In all cases, however, $k=4$ substantially outperforms $k=1$: on GPT-2 Large, adding data providers improves from 71.5\% to 80.5\% on MNLI and from 69.8\% to 75.0\% F1 on SQuAD; on Llama-3.2-1B, the gains are from 72.0\% to 83.0\% and from 71.0\% to 76.1\% F1. This gap is expected because each data provider only sees a quarter of the data, so even the union of their adapters does not perfectly match the representations learned from the full corpus. Nevertheless, on all datasets, $k=4$ is much closer to the upper bound than any single data provider and consistently benefits from adding more data providers, demonstrating that aggregation is a practical way to utilize professional knowledge from multiple enclaves.

We also evaluate how AlignAgg scales as the number of authorized data providers grows. Starting from the four trained adapters above, we synthetically scale to $k \in \{1,2,4,8,16,32\}$ logical enclaves by replicating adapters and assigning each replica to its own enclave. For each $k$, we authorize all enclaves for a single client and run AlignAgg so that every request uses all $k$ data providers. Table~\ref{tab:rq3-scale} reports the resulting per-request latency. Latency grows sublinearly with $k$: for GPT-2 Large, increasing from $k{=}1$ to $k{=}32$ enclaves raises latency from 12.0 sec to 32.4 sec (2.7$\times$); for Llama-3.2-1B, latency goes from 8.3 sec to 27.7 sec (3.3$\times$). Relative to a two-enclave configuration, even $k{=}32$ enclaves incur only a 2.1$\times$ slowdown on GPT-2 Large and a 2.2$\times$ slowdown on Llama-3.2-1B, far from the 32 $\times$ slowdown that a naive multi-model design (one model per provider) would suffer at $k{=}32$.

The result reflects how SwiftSched schedules split execution. As $k$ increases, AlignAgg must send activations to more enclaves and aggregate more enclaves, so communication volume and enclave-side work both grow and latency increases. However, SwiftSched keeps the number of GPU–TEE round-trips per token fixed and processes enclave calls in parallel, overlapping enclave computation with GPU decoding. Once a moderate number of enclaves is active (around 16 in our measurements), the GPU and the batched round-trips form the critical path, and additional enclaves mostly add constant-factor overhead rather than fully sequential work. This is why the latency increase from $k{=}16$ to $k{=}32$ is small (about 4\% for GPT-2 Large and 6\% for Llama-3.2-1B) and the curve begins to flatten. Overall, AlignAgg can benefit from multiple sources of professional knowledge while keeping latency manageable even as the number of contributing enclaves scales to dozens of data providers.

\subsection{RQ4: Effectiveness of SwiftSched}

\textbf{Baseline configurations.} We compare three runtime schedulers:
\textit{(1) PKUS-cpu} that runs the entire PKUS system on CPU without GPU acceleration. This represents the pure TEE-based execution where both the backbone and adapters execute inside enclaves.
\textit{(2) PKUS-pipeline} that enables asynchronous GPU-enclave communication to overlap computation, but lacks batched boundary crossing and adaptive scheduling.
\textit{(3) PKUS (SwiftSched)} denotes the full system combining pipeline parallelism, batched crossing, and adaptive work-stealing scheduling.

\begin{table}[t]
  \centering
  \caption{Per-request latency (sec) under different schedulers for single-data provider inference. Lower is better.}
  \label{tab:rq4-single}
  \small
  \begin{tabular}{lccc}
    \toprule
    Configuration & PKUS-cpu & PKUS-pipeline & PKUS \\
    \midrule
    Latency (sec)  & 83.8     & 124.2         & 10.4 \\
    Slowdown       & 8.1×     & 11.9×         & 1.0× \\
    \bottomrule
  \end{tabular}
\end{table}

Table~\ref{tab:rq4-single} reports per-request latency for end-to-end inference on extractive QA tasks. PKUS-cpu, which runs the entire model on CPU inside TEEs without GPU acceleration, incurs \textbf{8.1×} slowdown (83.8 sec per request) compared to PKUS. Interestingly, PKUS-pipeline exhibits even higher overhead at \textbf{11.9×} slowdown (124.2 sec), despite enabling GPU acceleration for the backbone. This performance degradation occurs because the fine-grained cross-boundary communication overhead dominates and overshadows any computational benefits from GPU acceleration. Without batching mechanisms, PKUS-pipeline must issue a separate message for each injection site, resulting in frequent context switches and serialization costs that compound with the already-limited GPU utilization under naive pipelining.

In contrast, PKUS (SwiftSched) achieves the lowest latency at 10.4 sec by addressing these bottlenecks through three complementary mechanisms. First, batched boundary crossing groups multiple injection sites into single messages, dramatically reducing the number of cross-boundary interactions. Second, adaptive work-stealing scheduling ensures balanced load distribution across enclaves and minimizes idle time. Third, the combination of pipeline parallelism with batching allows GPU computation to proceed while enclaves process aggregated requests, effectively hiding communication latency behind useful work. These results demonstrate that simply enabling GPU acceleration is insufficient—effective split execution requires co-designed communication optimization to realize performance gains.

The surprising performance degradation of PKUS-pipeline compared to PKUS-cpu highlights a critical insight: GPU acceleration alone is insufficient without co-designed communication optimization. PKUS successfully combines GPU throughput with effective cross-boundary batching and scheduling to achieve practical performance, reducing per-request latency by \textbf{8-12×} compared to baselines while preserving PKUS's strong isolation guarantees. This confirms that SwiftSched's mechanisms are essential for enabling practical deployment of trustworthy and controllable professional-knowledge utilization.

\subsection{RQ5: Security Analysis}

We analyze PKUS's security properties through qualitative comparison with MPC and HE, focusing on four dimensions: confidentiality, runtime access control, revocability and lifecycle management.

\textbf{Confidentiality.} 
PKUS achieves strong confidentiality through hardware-rooted TEE isolation, ensuring that adapters and computations remain inaccessible to the cloud operator. It operates on a single commodity TEE-enabled cloud server and natively supports arbitrary operators and full-featured neural architectures, therefore requiring no model-specific redesign. In contrast, MPC provides information-theoretic confidentiality only under the assumption of non-colluding servers, necessitating complex multi-party coordination, while HE relies on the hardness of lattice-based problems but demands extensive algorithmic adaptation to approximate non-polynomial operations in LLMs and suffers severe performance overhead. 

\textbf{Runtime access control.} 
PKUS enables fine-grained, policy-driven runtime access control via AegisProto, supporting role-based, context-aware, and time-bound restrictions. All policy checks execute inside the enclave, where queries and adapters are temporarily decrypted but isolated from the cloud operator, enabling semantic inspection without leakage. In contrast, MPC cannot enforce context-dependent policies without revealing query semantics to all servers, as partial model execution is required. Worse yet, HE provides no runtime policy enforcement at all, since encrypted computation prevents any plaintext observation. As a result, dynamic policy checks like token validation must be hard-coded into static homomorphic circuits, which is an infeasible approach for real-world, evolving requirements.

\textbf{Knowledge revocability.} 
PKUS enables fine-grained knowledge revocation without affecting other users or the backbone model. AegisProto then verifies the request, destroys the enclave, and securely erases the adapter in seconds, leaving other adapters in the same task unaffected. In contrast, both MPC and HE lack support for fine-grained revocation. Excluding a single adapter requires global recomputation in MPC or full decryption and re-encryption in HE, both of which are computationally prohibitive for LLMs and disrupt other participants as well.

\textbf{Lifecycle management.} 
All lifecycle operations in PKUS (adapter deployment, policy updates, revocations) are cryptographically signed by the enclave hardware, and bound to a remote attestation report. This ensures tamper evidence and non-repudiation for every operation.
Any authorized participant can independently verify and audit the complete operation history by inspecting these attestation reports. 
In contrast, neither MPC nor HE supports operation logging or external verification. MPC reveals no global view of execution beyond individual inputs, and HE processes ciphertexts without observing operational semantics. Consequently, it is impossible to determine who performed the operation, when it happened, or whether it was legitimate, making both approaches fundamentally incompatible with accountability requirements.

In summary, PKUS achieves confidentiality comparable to MPC and HE under the TEE trust model, while providing superior support for policy-based access control, fine-grained knowledge revocability, and auditable lifecycle management, which are properties that are difficult or infeasible with cryptographic approaches.

\section{Discussion and Related Work}

In prior works, HE and MPC-based systems~\cite{zeng2025mpcache,lu2023bumblebee,moon2025thor} show that inference and training on encrypted or secret-shared data are possible, but their computation and communication costs grow quickly with model size and depth, making them hard to scale to autoregressive LLMs and interactive workloads~\cite{9352960,deencryptedllm,zeng2025mpcache}. TEE-based systems instead execute models or slices of models inside enclaves~\cite{hashemi2021darknight,sun2020shadownet,hunt2018chiron,bai2025phantom,huang2022stamp,asvadishirehjini2022ginn,mo2020darknetz,tramer2019slalom}. One recent work compresses sensitive slices of a DNN into TEEs~\cite{li2025teeslice,zhang2024no}. However, these techniques are not optimized for LLMs and do not target scenarios that utilizing profesional knowledge from multiple data providers. PKUS treats adapters as the primary protection unit, protecting them in per-data provider enclaves, specifically to keep enclave computation small split execution practical for LLM inference.
\section{Conclusion}

This paper presents PKUS, a practical system enabling trustworthy protection and controllable utilization of professional knowledge in LLMs. By executing compact LoRA adapters within TEEs while keeping backbone models on GPUs, PKUS achieves strong confidentiality and auditable control without sacrificing utility or efficiency. Our evaluation shows PKUS matches the utility of full fine-tuning across diverse benchmarks, while reducing adapter parameters by 63-80\% and achieving 8-12$\times$ latency improvement over naive TEE baselines. PKUS provides a foundation for organizations to confidently share and monetize proprietary expertise while maintaining control over sensitive intellectual property.

\bibliographystyle{plain}
\bibliography{sample}


\end{document}